\documentstyle[aps,prl,amsfonts,twocolumn]{revtex}
\input epsf
\draft
\begin{document}
\title{Critical behaviour of three-dimensional Ising ferromagnets at imperfect surfaces:
Bounds on the surface critical exponent $\beta_1$}
\author{H.\ W.\ Diehl} 
\address{Fachbereich Physik, Universit\"at - Gesamthochschule Essen,\\
D-45117 Essen, Federal Republic of Germany}
\date{\today}
\maketitle
\begin{abstract}
The critical behaviour of three-dimensional semi-infinite 
Ising ferromagnets at planar surfaces with
(i) random surface-bond disorder or
(ii) a terrace of monatomic height and macroscopic size is considered.
The Griffiths-Kelly-Sherman correlation inequalities are shown to impose
constraints on the order-parameter density at the
surface, which yield upper and lower bounds 
for the surface critical exponent $\beta_1$.
If the surface bonds do not exceed the threshold for
supercritical enhancement of the pure system,
these bounds force $\beta_1$ to take the value
$\beta_1^{\text{ord}}$ of the latter system's
ordinary transition.
This explains the robustness of $\beta_1^{\text{ord}}$
to such surface imperfections observed in recent
Monte Carlo simulations.
\end{abstract}
\narrowtext
%
In a recent paper Pleimling and Selke (PS)  \cite{PS98}
reported the results of a detailed Monte
Carlo analysis of the effects
of two types of surface imperfections
on the surface critical behaviour 
of $d=3$ dimensional semi-infinite Ising models with
planar surfaces and
ferromagnetic nearest-neighbour (NN)
interactions: (i) random surface-bond disorder and
(ii) a terrace of monatomic height and macroscopic size on the surface.
For type (i), both the ordinary and special transitions were
studied. They found that the asymptotic temperature dependence
of the disorder-averaged surface magnetization
on approaching the bulk critical temperature $T_c$ from below
could be represented by a power law $\sim |\tau|^{\beta_1}$
with $\tau\equiv (T-T_c)/T_c$, where $\beta_1$  agreed,
within the available numerical accuracy, with
the respective values $\beta_1^{\text{ord}}\simeq 0.8$
and $\beta_1^{\text{sp}}\simeq 0.2$ of the pure system's
ordinary and special transitions. For type (ii), where the interaction
constants were chosen such that only an ordinary transition
could occur, the same value $\beta_1^{\text{ord}}$ of the perfect
system was found for $\beta_1$.

Their findings for the case of (i)
are in conformity with the
relevance/irrelevance criteria of Diehl and N{\"u}sser \cite{DN90a,Har74}
according to which the pure system's surface critical behaviour
should be expected to be stable
or unstable with respect to short-range correlated random
surface-bond disorder depending on whether
the surface specific heat $C_{11}$ \cite{Die86a}
of the pure system remains finite
or diverges at the transition. It is fairly well
established \cite{DD83b,DDE83} that $C_{11}$ approaches a finite constant
at the ordinary transition, but has a leading thermal singularity
$\sim|\tau|^{(d-1)\nu -2\Phi}$ at the special transition,
where $\Phi$ is the surface crossover exponent.
In the latter case, the condition for irrelevance,
$\Phi<(d-1)\nu/2$, reduces to
\begin{equation}\label{irrelcon}
\Phi<\nu 
\end{equation}
in $d=3$ bulk dimensions.
Since various Monte Carlo simulations
\cite{LB90a,RDW92,RDWW93}
(though not all
\cite{vrf})
and renewed field-theory
estimates \cite{DS94}
suggest a value of
$\Phi$ between $0.5$ and $0.6$, definitely
smaller than the accepted value $0.63$ of $\nu$ for $d=3$,
one may be quite confident that the
condition (\ref{irrelcon}) holds. Thus short-range correlated
surface-bond disorder should be irrelevant
in the renormalization-group sense at both transitions.

Irrelevance criteria of the above Harris type \cite{DN90a,Har74}
seem to work quite well in practice. Yet,
from a mathematical point of view,
they are rather weak because they are
nothing but a necessary (though not sufficient)
condition for stability of the pure system's critical behaviour.

In this note, I shall employ the Griffiths-Kelly-Sherman (GKS)
inequalities \cite{Gri67a}
to obtain upper and lower bounds on the
surface magnetization densities
of both types of imperfect systems, bounds
that are given by surface magnetizations of analogous
systems without such imperfections.
Their known asymptotic temperature dependence
near $T_c$ will then be exploited to obtain restrictions
on the surface critical behaviour of the imperfect systems
considered. For some cases of interest studied by PS \cite{PS98},
the equality $\beta_1=\beta_1^{\text{ord}}$ will be rigorously
established.
 
Following these authors, let us consider an Ising model
with ferromagnetic NN interactions on a simple cubic lattice of size
$L_x\times L_y\times L_z$. Periodic boundary conditions will be chosen
along two principal axes (the $x$ and $y$ directions),
and free boundary conditions along the third one (the $z$ direction),
so that the surface consists of the top layer at $z=1$ and the bottom
layer at $z=L_z$. Associated with each pair of spins on
NN sites $i$ and $j$ is an interaction constant $J(i,j)>0$,
which we assume to have
the same value $J$ whenever $i$ or $j$ (or both) belong to
layers with $1<z<L_z$.

In the case of surface-bond disorder,
which we consider first, the
$J(i,j)\equiv J^{\text{(s)}}(i,j)$ of
all NN pairs of surface sites are
independent random variables. The probability density
$P(J_1)$ of any one of these will be assumed to have support
only in the interval $[J_1^<,J_1^>]$ (with $J_1^>>J_1^<>0$).
This is in conformity with, but less restrictive than, PS's
assumption that $J_1$ takes just two values $J_1^<$ and $J_1^>$,
either one with probability $1/2$. We will also assume that all
(bulk and surface) spins are exposed to the same magnetic field $H>0$,
whose limit $H\to 0^+$ will be taken after the thermodynamic limit
has been performed.

Let $K\equiv J/k_BT$ and $h\equiv H/J$. Define
$\bbox{r}^{(\text{s})}$ to be
the set of all dimensionless surface
coupling constants $J^{(\text{s})}(i,j)/J$. Let
$m(i;K,\bbox{r}^{(\text{s})},h)\equiv
\langle s_i\rangle$ be the thermal average of a spin at site $i$
for a given disorder configuration $\bbox{r}^{(\text{s})}$,
and denote
the corresponding quantity of the perfect system with uniform
NN surface coupling $J_1=rJ$ as
$m(i;K,r,h)$. Since all interactions are ferromagnetic,
the GKS inequalities \cite{Gri67a} are valid. Averages
of products of spin variables are monotone non-decreasing
functions of all variables $J(i,j)$ and $H$. Hence, for finite $L_x,\,
L_y$, and $L_z$, $m(i;K,\bbox{r}^{(\text{s})},h)$
is bounded by $m(i;K,r^<,h)$ from below and by
$m(i;K,r^>,h)$ from above.
We choose $i\equiv i_s$ to be a surface site,
take the thermodynamic limit
(first) and then let $H\to 0^+$.
The bounds converge towards
the respective values of $m_1(K,r,0^+)$,
the spontaneous magnetization of the surface
layers per site, for $r=r^<$ and $r^>$.
Thus we obtain
\begin{equation}\label{Grifineq}
m_1(K,r^<,0^+)\le m(i_s;K,\bbox{r}^{(\text{s})},0^+)\le
m_1(K,r^>,0^+)\;.
\end{equation}

The following limiting forms of $m_1$ are well
established \cite{PS98,Die86a,LB90a,rigres,BD94}:
\begin{equation}\label{limform}
m_1=\cases{C_1|\tau|^{\beta_1^{\text{ord}}}[1+o(\tau)]&
as $\tau\to 0^-$ at fixed  $r<r_c$,\cr
C_1'|\tau|^{\beta_1^{\text{sp}}}[1+o(\tau)]
&as $\tau\to 0^-$ at fixed  $r=r_c$,\cr
m_{1c}+O(\tau)
&as $\tau\to 0^\pm$ at fixed $r>r_c$,}
\end{equation}
where $r_c\simeq 1.50 $ \cite{LB90a} is the critical value associated with
the special transition.
The quantities $m_{1c}>0$, $C_1$, and $C'_1$ are
nonuniversal, whence the first two depend on $r$.

Consider first the case $r^><r_c$. Let $C^<$
and $C^>$ be the values of $C_1$ for $r=r^<$
and $r=r^>$, respectively. (These satisfy
$0<C^<\le C^><\infty$ provided $0<J<\infty$
and $0<J_1^<\le J_1^><\infty$.) It follows that
there exists a number $\epsilon >0$ independent
of the disorder configuration $\bbox{r}^{\text{(s)}}$
such that
\begin{equation}\label{ineq}
C^>\le m[i_s;K(\tau),\bbox{r}^{\text{(s)}},0^+]\,
\,|\tau|^{-\beta_1^{\text{ord}}}\le C^>
\end{equation}
whenever $-\epsilon <\tau <0$. We denote the
average of a quantity $Q$ over all choices of the random
variables $\bbox{r}^{\text{(s)}}$ as $\overline{Q}$.
Upon averaging $m(i_s;.)$ to obtain the disorder-averaged
surface magnetization $\overline{m_1}$, we see that
the inequality (\ref{ineq}) holds for
$\overline{m_1}\,|\tau|^{-\beta_1^{\text{ord}}}$
as well.
An elementary consequence is: If $\overline{m_1}$
has a well-defined critical exponent $\beta_1^{\text{dis}}$
in the sense that \cite{Sta71}
\begin{equation}\label{defce}
\beta_1^{\text{dis}}=\lim_{\tau\to 0^-}
\frac{\ln \overline{m_1}(\tau)}{\ln \tau}
\end{equation}
exists, then we have
\begin{equation}\label{ordeq}
\beta_1^{\text{dis}}=\beta_1^{\text{ord}}\,.
\end{equation}

Two further implications of (\ref{ineq}) are worth
mentioning. First, if a surface critical
exponent $\tilde\beta_1^{\text{dis}}$ can be defined
via the analog of (\ref{defce}) for the most probable
value of $m(i_s;.)$ \cite{comment}, then it must have the same
value $\beta_1^{\text{ord}}$.
Second, the inequality
(\ref{ineq}) also rules out a limiting $\tau$
dependence of the form
$\sim |\tau|^{\beta_1}\,|\ln|\tau||^\varphi$
(standard logarithmic corrections)
for $\overline{m_1}$ and the most probable
value of $m(i_s;.)$.

Consider next the case $r^>=r_c$. Let us again make the assumption that
the limit (\ref{defce}) or the analogous one defining
$\tilde\beta_1^{\text{dis}}$ exist. Then the inequalities
\begin{equation}\label{betadisineq}
\beta_1^{\text{sp}}\le \beta_1^{\text{dis}}\le \beta_1^{\text{ord}}
\end{equation}
and their analogs for $\tilde\beta_1^{\text{dis}}$ can be deduced
from (\ref{ineq}). (Cf.\ Lemma 3 of \cite{Sta71}.)

The same reasoning applied in the case $r^>>r_c$ shows that
$\beta_1^{\text{dis}}$ or $\tilde\beta_1^{\text{dis}}$
must obey the relations
\begin{equation}
0\le \beta_1^{\text{dis}}\le \beta_1^{\text{ord}}
\end{equation}
whenever the limits (\ref{defce}) through which we defined them exist.

Likewise in the case $r^<=r_c$,
the possible values of $\beta_1^{\text{dis}}$ or
$\tilde\beta_1^{\text{dis}}$ are restricted by
\begin{equation}\label{bsp}
0\le \beta_1^{\text{dis}}\le \beta_1^{\text{sp}}
\end{equation}
at transitions at which $\overline{m_1}$ or the most
probable value of $m(i_s;.)$  \cite{comment}
approach zero, respectively.
On the other hand, it should be recalled
that the surface critical exponent
$\beta_1^{\text{ex}}$ of the pure system's extraordinary transition
requires a definition other than (\ref{defce}): One must
subtract a regular background contribution $m_1^{\text{reg}}$
from $m_1$ and define $\beta_1^{\text{ex}}$
through the limiting behaviour
$m_1-m_1^{\text{reg}}\sim |\tau|^{\beta_1^{\text{ex}}}$.
For transitions of the impure systems at which $\overline{m_1}$
approaches a constant $\ne 0$, it would also
not make much sense to define $\beta_1^{\text{dis}}$
via (\ref{defce}). Of course, for surface critical exponents
$\beta_1^{\text{dis}}$ not given by (\ref{defce}), the above
bounds do {\em not\/} apply.
This means that they cannot be utilized
to draw conclusions about the
surface critical exponent $\beta^{\text{ex}}_1$
of the impure system's extraordinary transition.
However, for a special transition of the impure system with
$\overline{m_1}(\tau=0)=0$, the inequalities (\ref{bsp}) hold.

The inequality (\ref{Grifineq}) rules out that the impure system
has an ordered surface phase for $T>T_c$ whenever $r^>\le r_c$.
In order that the impure system can have an extraordinary or special transition,
the distribution $P(J_1)$ of the surface couplings typically will have to extend
beyond the critical-enhancement threshold $r_cJ$ of the pure system.
But even if $r^>>r_c$, an ordinary transition may still occur if
the surface bonds `on average' are not sufficiently enhanced
(cf.\ \cite{PS98}). However, if $P(J_1)$ extends beyond $r_cJ$,
then disorder configurations for which
macroscopically large surface regions have
the same supercritical value ($>r_cJ$) of $J_1$ occur with finite
probability. This happens even if the impure system (for a typical realization
of disorder) undergoes an ordinary transition, albeit with exponentially
small probability. By analogy with the bulk case \cite{Gri69},
I expect surface quantities like
$\overline{m_1}$ and the disorder-averaged surface free energy
to be {\em non-analytic\/} functions
of the surface magnetic field $H_1$ at $H_1=0$
for  temperatures between the bulk critical temperature $T_c$ and the
temperature $T_s(r^>)>T_c$ at which
the semi-infinite pure system with homogeneous
surface coupling $J_1=r^>J$ undergoes a transition to a surface-ordered,
bulk-disordered phase. That is, they should display
{\em Griffiths singularities\/} \cite{Gri69}, a problem
on which we will not embark further here.

Turning now to the case of surfaces with a terrace,
we start from a pure Ising model of the sort considered above.
Just as PS, we assume that {\it all\/} NN couplings $J(i,j)$
(including those between surface sites)
have the same value $J$. Let us denote thermal averages
pertaining to this system by a superscript $[I]$, writing, e.g.,
$m^{[I]}(i;K)=\langle s_i\rangle^{[I]}$.
We consider another system, $[II]$, which differs from $[I]$ through
the addition of a zeroth layer at $z=0$ whose spins are assumed
to interact among themselves and with the spins in the  $z=1$ layer
via NN interaction constants $J_1$ and $J$, respectively.
To obtain a system with a terrace, $[T]$, we choose a subregion
of the zeroth layer (the terrace) and remove all those NN bonds
$J$ and $J_1$ that are connected to lattice sites of this layer
outside the terrace region. PS considered a strip-like terrace
of size $(L_x/2)\times L_y$, and assumed that $J_1=J$.
For our considerations, the precise form
and  size of the terrace region
will not be important. (One could even
assume that an arbitrary subset of the spins in the zeroth
layer are decoupled from the rest of the system.)

Let $i_1$ be an arbitrary lattice site in the $z=1$ layer. Since
the systems $[I]$, $[T]$, and $[II]$ differ by the addition of ferromagnetic
interactions, we have from
the GKS inequalities,
\begin{equation}
m^{[I]}(i_1;K,h)\le m^{[T]}(i_1;K,r,h)\le
m^{[II]}(i_1;K,r,h)
\end{equation}
where, as before, $h=H/J>0$ is a uniform magnetic field
and  $r=J_1/J$. In the thermodynamic limit $L_x,\,L_y,\,
L_z\to\infty$, the lower  and upper bounds converge towards
$m_1(K,h)$, the magnetization per site of the topmost layer, and
to $m_2(K,r,h)$, the magnetization per site of the layer underneath the
topmost layer, respectively.
If we assume that $r<r_c$ (subcritical surface enhancement)
and take the limit $h\to 0^+$, then the limiting form shown in the first
line of (\ref{limform}) applies to both $m_1$ and $m_2$  (with different
values of $C_1$). As a straightforward consequence we find
that the surface critical exponent $\beta_1$ of
$m^{[T]}(i_1;K,r,0^+)$ (for an arbitrary site $i_1$ with $z=0$)
strictly satisfies  $\beta_1=\beta_1^{\text{ord}}$.

It evident that the same reasoning can be applied to the analogous
two-dimensional model with a terrace to conclude that $\beta_1$
takes the exactly known value $\beta_1^{\text{ord}}=1/2$.
Likewise, the inequality (\ref{ineq}) and the
result (\ref{ordeq}) carry over to the two-dimensional
case, giving $\beta_1^{\text{dis}}=1/2$ for all values of $r<\infty$, since
$r_c=\infty$ for $d=2$. Note also that the inequality (\ref{ineq})
excludes the possibility of an asymptotic temperature dependence of the
form  $\overline{m_1}\approx \text{const } |\tau|^{1/2}|\ln|\tau||^p$
(i.e., of logarithmic correction factors). This is because it is known
for the pure case that
no such logarithmic corrections appear in the limiting form of $m_1$.

Results of Monte Carlo simulations on the
surface critical behaviour of  two-dimensional Ising models
with bond disorder have been reported 
in two recent papers \cite{SSLI97}. 
However, in this work
random bond disorder was assumed to be present both
in the bulk and at the surface, a case not captured by our reasoning.
Nevertheless, $\overline{m_1}$ was found to behave as $|\tau|^{1/2}$,
apparently without logarithmic corrections, even though the presence of
such a correction could be detected in the limiting form
of the disorder-averaged bulk
order parameter.

I am indebted to W.\ Selke for informing me about the work
\cite{PS98} prior to publication, and to him, Joachim Krug,
and Kay Wiese for
a critical reading of the manuscript.
This work has been supported by the Deutsche Forschungsgemeinschaft
through the Leibniz program.


\begin{references}
\bibitem{comment} Quantities like $m_1\equiv \sum_{i_s}\langle s_{i_s}\rangle/\sum_{i_s}1$, the total magnetization of the surface
 per spin, may be expected to be self-averaging
in the thermodynamic limit. Hence the value $m_1$ takes  for any
choice of the random variables $\bbox{r}^{\text{(s)}}$ that occurs with
reasonable probability should agree with $\overline{m_1}$.
Once one accepts this, the distinction between
$\beta^{\text{dis}}_1$ and $\tilde\beta^{\text{dis}}_1$ becomes
unnecessary. It is made here because the assumption of self-averaging
is neither required nor used in the derivation
of the given inequalities,
and not to suggest a possible breakdown of self-averaging.
\end{references}
\end{document}